\magnification=1200\baselineskip18pt

\def\a{\alpha}

\def\k{\kappa}\def\l{\lambda}\def\m{\mu}\def\n{\nu}\def
\p{\pi}\def\r{\rho}\def\s{\sigma}
\def\y{\eta}\def\x{\xi}

\def\L{\Lambda}
\def\O{\Omega}

\def\de{\partial}
\def\id{\equiv}\def\mo{{-1}}\def\ha{{1\over 2}}

\def\mn{{\mu\nu}}

\def\pun{\cdotp\!}

\def\coo{coordinates }

\def\cosc{cosmological constant }

\def\pb{Poisson brackets }

\def\ads{anti-de Sitter }
\def\poi{Poincar\'e }
\def\des{de Sitter }

\def\SR{special relativity }

\def\tls{transformation laws }

\def\section#1{\bigskip\noindent{\bf#1}\smallskip}

\def\PL#1{Phys.\ Lett.\ {\bf#1}}
\def\PRL#1{Phys.\ Rev.\ Lett.\ {\bf#1}}
\def\PR#1{Phys.\ Rev.\ {\bf#1}}\def\CQG#1{Class.\ Quantum Grav.\ {\bf#1}}

 \def\IJMP#1{Int.\ J. Mod.\ Phys.\ {\bf #1}}
\def\MPL#1{Mod.\ Phys.\ Lett.\ {\bf #1}} 

\def\AoP#1{Ann.\ Phys.\ {\bf#1}}
\def\hep#1{{\tt hep-th/#1}}\def\arx#1{{\tt arXiv:#1}}

\def\ref#1{\medskip\everypar={\hangindent 2\parindent}#1}
\def\beginref{\begingroup
\bigskip
\centerline{\bf References}
\nobreak\noindent}
\def\endref{\par\endgroup}

\def\lox{1-\L\O\,x^2p^2}\def\op{1-\O\,p^2}\def\lx{1-\L\,x^2}


{\nopagenumbers
\line{\hfil August 2008}
\vskip40pt
\centerline{\bf Doubly special relativity and translation invariance}
\vskip60pt
\centerline{
{\bf S. Mignemi}\footnote{$^\ddagger$}{e-mail:
smignemi@unica.it}}
\vskip10pt
\centerline {Dipartimento di Matematica, Universit\`a di Cagliari}
\centerline{viale Merello 92, 09123 Cagliari, Italy}
\centerline{and INFN, Sezione di Cagliari}
\vskip80pt
\centerline{\bf Abstract}

\vskip10pt
{\noindent
We propose a new interpretation of doubly special relativity (DSR) based on the distinction
between the momentum and the translation generators in its phase space realization.
We also argue that the implementation of DSR theories does not necessarily require a
deformation of the Lorentz symmetry, but only of the translation invariance.
}
\vskip100pt\
P.A.C.S. Numbers: 11.30.Cp, 03.30.+p
\vfil\eject}

\section{Introduction}
In recent years, the idea that special relativity should be modified for energies
close to the Planck scale $\k$, in such a way that $\k$ becomes an observer-independent
parameter of the theory like the speed of light, has been extensively debated [1-4].
This hypothesis is motivated by the consideration that the Planck energy sets a limit
above which quantum gravity effects become important, and its value should therefore not
depend on the specific observer, as would be the case in special relativity.
Of course, this postulate must be implemented in such a way that the principle of
relativity, i.e.\ the equivalence of all inertial observers, be still valid.
The theory based on these assumptions has been named doubly special relativity (DSR)
[1].

DSR models are realized by deforming the \poi invariance of special relativity.
Their main physical consequences are the modification of the dispersion relations
of elementary particles and the existence of a nonlinear addition law for the
momenta. In particular, one is lead to identify $\k$ with a maximal value of the
energy or the momentum for elementary particles.

These nontrivial effects have been used to derive experimentally verifiable predictions,
for example to explain the observed threshold anomalies in ultrahigh-energy cosmic rays
[5].
Another natural, although not necessary, consequence of the formalism is the
noncommutativity of the spacetime geometry.
In particular, DSR fits very well in the formalism of $\k$-\poi algebras [6], that
postulates a quantum deformation of the \poi group acting on noncommutative spacetime
[4], although the two theories cannot be considered equivalent [7].
One drawback of DSR is however that the deformation of \SR resulting from its
postulates is not unique, and several inequivalent models can be constructed.

DSR is usually associated with the deformation of the Lorentz symmetry.
In this paper we wish to point out that its really distinguishing feature is not
the deformation of the Lorentz symmetry, but rather that of the translation symmetry.
As mentioned before, the main phenomenological consequences of DSR are a deformation
of the addition law of momenta and of the dispersion law of the elementary particles.
These clearly depend only on the nontrivial action of translations generators on
momenta in phase space. In fact, the
deformed dispersion relation is given by the Casimir invariant of the translations.

The lack of the necessity of deforming the Lorentz invariance is clearly illustrated
for example by the Snyder realization of DSR [8,9].
The Snyder model was originally proposed [8] in order to show the possibility of
introducing a noncommutative spacetime without breaking the Lorentz symmetry.
It was then observed that it can be interpreted as a DSR model [9], but the physical
implications of this fact were not further investigated.
In the original formulation of the model [8], the \poi invariance is realized canonically,
but, as we shall show, it is possible to give an alternative interpretation in terms of
DSR.

Another important point we wish to stress is that, from an algebraic point of
view, the realization
of DSR does not necessarily require a deformation of the \poi group, as in the $\k$-\poi
formalism, but can be carried out in a classical framework through a nonlinear action
of the \poi group on phase space. This interpretation of DSR is close in the spirit
to the proposal of [3] and has been stressed especially in [10].
In essence, at the classical level one can deform the generators of the \poi
group in such a way that obey the standard \poi algebra but nevertheless act nontrivially
on phase space variables.

In the present paper we show that these ideas can be implemented in a natural way if,
in analogy with what happens in curved space, one distinguishes the translation generators
from the canonical momenta.
This observation also clarifies the physical origin of the composition law of momenta
proposed in ref.\ [11] and usually adopted in the DSR literature, whose interpretation was
rather obscure. In fact, in that paper the addition of momenta was obtained through the
introduction of unphysical auxiliary variables, that in our interpretation are identified
with the generators of the translation symmetry.

To illustrate these considerations, we discuss the Snyder model from a DSR point of view.
The same formalism can of course be applied to more traditional DSR models, where also
the action of the Lorentz group is deformed.
\section{The model}
Let us start by considering the classical action of the \poi algebra on the phase space of
special relativity\footnote{$^1$}{We adopt the notations $\m=0,\dots,3$, $A=0,\dots,4$,
$i=1,\dots,3$, $x\pun p\id x_\m p_\m$.}.
The \poi algebra is spanned by the Lorentz generators $J_\mn$ and the translation
generators $T_\m$, obeying Poisson brackets
$$\eqalignno{
&\{J_\mn,J_{\r\s}\}=\y_{\n\s}J_{\m\r}-\y_{\n\r}J_{\m\s}+\y_{\m\r}J_{\n\s}-
\y_{\m\s}J_{\n\r},&\cr
&\quad\{J_\mn,T_\l\}=\y_{\m\l}T_\n-\y_{\n\l}T_\m,\qquad\{T_\m,T_\n\}=0.&(1)}$$
Its realization in canonical phase space, with \pb
$$\{x_\m,x_\n\}=\{p_\m,p_\n\}=0,\qquad\{x_\m,p_\n\}=\y_\mn,\eqno(2)$$
is obtained through the identification
$$J_\mn=x_\m p_\n-x_\n p_\m,\qquad T_\m=p_\m,\eqno(3)$$
which yields the \tls for the phase space coordinates,
$$\eqalignno{&\{J_{\mn},x_\l\}=\y_{\m\l}x_\n-\y_{\n\l}x_\m,\qquad
\{J_{\mn},p_\l\}=\y_{\m\l}p_\n-\y_{\n\l}p_\m,&\cr
&\qquad\qquad\quad\{T_\m,x_\n\}=\y_\mn,\qquad\{T_\m,p_\n\}=0.&(4)}$$

This formalism can be easily generalized to \des space with \cosc $\L$.
The \des algebra is given by
$$\eqalignno{
&\{J_\mn,J_{\r\s}\}=\y_{\n\s}J_{\m\r}-\y_{\n\r}J_{\m\s}+\y_{\m\r}J_{\n\s}-
\y_{\m\s}J_{\n\r},&\cr
&\{J_\mn,T_\l\}=\y_{\m\l}T_\n-\y_{\n\l}T_\m,\qquad\{T_\m,T_\n\}=-\L J_\mn.&(5)}$$
Usually, the algebra is realized in terms of the isometries of a hyperboloid of equation
$\x_A^2=-1/\L$, embedded in flat five-dimensional space with metric
$\y_{AB}={\rm diag}\,(1,-1,-1,-1,-1)$. In the following, we shall denote $\p_A$ the
momenta canonically conjugate to $\x_A$.
The Lorentz generators $J_\mn$ are identified with the corresponding generators of the
five-dimensional algebra, while the translation generators $T_\m$ are identified with
$\sqrt\L\,J_{4\m}$.

The realization of the \des algebra in four-dimensional phase space, with \pb (2),
depends on the specific \coo chosen on the hyperboloid.
In general, the Lorentz generators maintain the canonical form, $J_\mn=x_\m p_\n-x_\n p_\m$,
while the form of the translation generators depends on the parametrization of
the hyperboloid.
A convenient choice is given by Beltrami (projective) \coo [12,13],
$x_\m=\x_\m/\sqrt\L\,\x_4$, with canonically conjugate momentum $p_\m=\sqrt\L\,\x_4\,\p_\m$,
in terms of which the translation generators read
$$T_\m=p_\m-\L\,x\pun p\,x_\m.\eqno(6)$$
It is then evident that in \des spacetime the generators of the translation symmetry
cannot be identified with the canonical momenta $p_\m$.

Under translations, the Beltrami \coo $x_\m$ and $p_\m$ transform as
$$\{T_\m,x_\n\}=-\y_\mn+\L x_\m x_\n,\qquad\{T_\m,p_\n\}=-\L(x\pun p\,\y_\mn+x_\m p_\n).
\eqno(7)$$
Thus the conserved quantity is not the canonical momentum $p_\m$, as should be obvious since
the hamiltonian of a free particle in \des spacetime is position-dependent\footnote{$^2$}{For
example, in Beltrami \coo the hamiltonian of a free particle is given by
$H={1\over2}\ (1-\L x^2)[p^2-\L(x\pun p)^2]$.}, but the quantity associated
with the translation generator, given by (6).
The rules for the composition of momenta are dictated by the conservation of
$T_\m$ and not of $p_\m$.

\bigskip
We pass now to consider the case of DSR. As discussed previously, we adopt the point
of view that the symmetry algebra maintains its classical form, but its action
on phase space is nonlinear. Since our aim is to show the relevance of translation invariance,
we consider the specific example of the Snyder model, whose most noticeable feature is
that the Lorentz invariance is realized linearly in the standard way, but our
considerations can be extended to any other DSR model.

As it was shown in ref. [11], the \tls of any DSR model can be obtained by defining the
physical momentum $p_\m$ in terms of auxiliary variables $P_\m=U(p_\m)$ that satisfy
canonical transformation laws.
The deformed dispersion relation is then given by writing the classical relation for
the auxiliary variables, $P^2=m^2$ in terms of the physical variables $p_\m$.
Also the addition law of momenta is obtained by pulling back to the physical momenta
$p_\m$ the classical law for the variables $P_\m$.

We propose that, in analogy with the case of \des space, the generators of translations $T_\m$
should not be identified with the momenta $p_\m$, but rather with the auxiliary variables
$P_\m$.
This choice clarifies the physical significance of the auxiliary variables and of the
addition law for momenta proposed in [11].

In particular, in the case of the Snyder model, one chooses [13]
$$P_\m=U(p_\m)={p_\m\over\sqrt{1-\O p^2}},\eqno(8)$$
with inverse
$$p_\m=U^\mo(P_\m)={P_\m\over\sqrt{1+\O P^2}}.\eqno(9)$$

In a DSR interpretation, the Snyder model can then be characterized by the explicitly Lorentz
invariant deformed dispersion relation $p^2/(1-\O p^2)=m^2$, i.e.\ $p^2=m^2/(1+\O\,m^2)$, where
$\O=1/\k^2$ is the Planck area.
In this form, the dispersion relation looks like a redefinition of the mass (notice that
the dispersion relation for massless particles maintain its classical form), however some
nontrivial consequences follow$^{3,4}$\footnote{}{$^3$ The original paper [8] gives a
different interpretation of the physics, in which the classical dispersion relation
$p^2=m^2$ is maintained.

{$^4$} It may be interesting
to notice that the Snyder model can be derived from a 5-dimensional momentum space of \coo
$\p_A$, constrained by $\p_A^2=-1/\O$, in a way dual to that used for \des spacetime [9].}.

For example, for $\O>0$ the model admits a maximal mass $\k$, and is similar to other
DSR models, which admit a maximum value for the momentum or the energy of a particle.
For $\O<0$, instead, there is no limit value for the mass.
In the quantum theory however emerges the existence of a minimal value for the momentum, as
in the similar model discussed in [14].
In the following, we consider the case of positive $\O$.

From the structure of (8) it follows that the action of the Lorentz group on the momentum
variables is not affected, while that of translations is deformed. This illustrates
the fact that the most relevant characteristic for the implementation of DSR is
the deformation of the action of translations (and hence a modified composition law
of momenta) and not that of Lorentz transformations, as usually postulated.

In order to realize the model in spacetime, it is natural to introduce position
variables $x_\m$ that transform covariantly with respect to the momenta. These can be
defined as [13,15]
$$x_\m=\sqrt{1+\O P^2}\,X_\m,\eqno(10)$$
where $X_\m$ are the variables canonically conjugate to the $P_\m$.

With this definition, the \pb between the new phase space \coo are no longer
canonical, and the position space becomes noncommutative, realizing the proposal of
Snyder [8],
$$\{x_\m,x_\n\}=-\O(x_\m p_\n-x_\n p_\m),\qquad\{p_\m,p_\n\}=0,
\qquad\{x_\m,p_\n\}=\y_\mn-\O\,p_\m p_\n.\eqno(11)$$

In terms of the physical \coo $x_\m$ and $p_\m$, the generators of the \poi group read
$$J_\mn=x_\m p_\n-x_\n p_\m,\qquad T_\m=P_\m=p_\m/\sqrt{1-\O p^2}.\eqno(12)$$
The \tls of $x_\m$ and $p_\m$ under the Lorentz subalgebra maintain the canonical form,
while under translations become
$$\{T_\m,x_\n\}={\y_\mn\over\sqrt{1-\O p^2}},\qquad\{T_\m,p_\n\}=0.\eqno(13)$$
Therefore, the effect of the translations on the position coordinates becomes momentum
dependent and increases for near Planck-mass particles.

The sum of the momenta of two particles with momenta $p^{(1)}_\m$ and $p^{(2)}_\m$
is given in general by [3]
$$p^{(12)}_\m=U^\mo[U(p^{(1)}_\m)+U(p^{(2)}_\m)]$$
and in our case it can be readily obtained from (8) and (9),
$$p^{(12)}_\m={\sqrt{1-\O(p^{(2)}_\m)^2}\ p^{(1)}_\m+\sqrt{1-\O(p^{(2)}_\m)^2}\
p^{(2)}_\m\over\sqrt{1-\O^2(p^{(1)}_\m)^2(p^{(2)}_\m)^2+2\O p^{(1)}_\m\pun p^{(2)}_\m
\sqrt{(1-\O(p^{(1)}_\m)^2)(1-\O(p^{(2)}_\m)^2)}}}\ .\eqno(14)$$
Notice that this expression is nontrivial even for massless particles.
Using (14) one may calculate the effect of the deformed transformations
on the scattering of ultra-high-energy cosmic rays by the cosmic background radiation [5].
We shall not perform the calculation in detail, but a correction of the classical
threshold of electron production arises as in the other DSR models.

It is also possible to define a dynamics for the free particle, by introducing a
hamiltonian. This can be obtained simply substituting (8) into the classical hamiltonian,
$$H={P^2\over2}=\ha\ {p^2\over1-\O\,p^2}.\eqno(15)$$
The Hamilton equations can then be derived taking into account the symplectic structure
(11). They read
$$\dot x_\m={p_\m\over1-\O p^2},\qquad\dot p_\m=0.\eqno(16)$$

The 3-velocity of a free particle, defined as ${\bf v}_i=\dot x_i/\dot x_0$ maintains its
classical expression $p_i/p_0$ and cannot exceed the speed of light. The same expression
is obtained by the alternative definition ${\bf v}_i=\de p_0/\de p_i$.

Also a natural definition of the spacetime metric can be given [13], yielding
$ds^2=(1-\O p^2)dx^2$. As usual in DSR, the metric depends explicitly on the momentum
[15].

\section{Conclusion}

We have shown that DSR can be interpreted as a classical relativistic mechanic
model with nontrivial generators of translations,
and have discussed this point in the special case of the Snyder model.
Similar considerations can be applied to other DSR models. For example, the results
of [10] for the Maguejo-Smolin model [3] can be interpreted in this perspective.
The discussion of this specific model, which is Lorentz invariant, shows also that the
implementation of DSR does not necessarily require in general a deformation of the
Lorentz symmetry, but only of the translational one, contrary to the common view.
More general Lorentz-invariant DSR models can also be obtained starting from different
Lorentz-invariant deformations of the dispersion relation of elementary particles,
of the form $f(p^2)=m^2$.

Of course the crucial point for the physical interpretation is that the physically
measured variables should be identified with the canonical momenta. This would require
an operational definition of momentum measurements in DSR that to our knowledge is still
lacking.

We notice that our interpretation is not intended to solve the problems of DSR
such as the so-called soccer ball problem (i.e. the fact that if DSR had to hold also
for macroscopical objects, their momentum should not exceed the Planck scale), but just
to give a clearer interpretation of the formalism in classical terms, alternative for
example to the quantum-group based $\k$-\poi formalism. From the discussion above,
it should also be evident that DSR is no more equivalent to \SR than \des space is to
flat space.

\beginref
\ref [1] G. Amelino-Camelia, \PL{B510}, 255 (2001), \IJMP{D11}, 35 (2002).
\ref [2] J. Kowalski-Glikman, \PL{A286}, 391 (2001).
\ref [3] J. Magueijo and L. Smolin, \PRL{88}, 190403 (2002); \PR{D67}, 044017 (2003).
\ref [4] For a review, see J. Kowalski-Glikman, Lecture Notes in Physics {\bf 669}, 131
(2005).
\ref [5] G. Amelino-Camelia and T. Piran, \PL{B497}, 265 (2001); \PR{D64}, 036005 (2001).
\ref [6] S. Majid and H. Ruegg, \PL{B334}, 348 (1994);
J. Lukierski, H. Ruegg and W.J. Zakrzewski, \AoP{243}, 90 (1995).
\ref[7] G. Amelino-Camelia, J. Kowalski-Glikman, G. Mandanaci and A. Procaccini,
\IJMP{A20}, 6007 (2005).
\ref [8] H.S. Snyder, \PR{71}, 38 (1947).
\ref [9] J. Kowalski-Glikman and S. Nowak, \IJMP{D13}, 299 (2003).
\ref [10] S. Ghosh and P. Pal, \PR{D75}, 105021 (2007).
\ref [11] S. Judes and  M. Visser, \PR{D68}, 045001 (2003).
\ref [12] H.Y. Guo, C.G. Huang, Z. Xu and B. Zhou, \PL{A331}, 1 (2004).
\ref [13] S. Mignemi, \arx{0802.1129}.
\ref [14] A. Kempf, G. Mangano and R.B. Mann, \PR{D52}, 1108 (1995).
\ref [15] S. Mignemi, \PR{D68}, 065029 (2005).
\ref [16] J. Magueijo and L. Smolin, \CQG{21}, 1725 (2004).
\ref [17] J. Kowalski-Glikman and L. Smolin, \PR{D70}, 065020 (2004).
\end

\ref [9] J. Kowalski-Glikman, \MPL{A17}, 1 (2002).
\ref [10] A. Granik, \hep{0207113}.

\ref [15] J. Kowalski-Glikman, \PL{B547}, 291 (2002);
J. Kowalski-Glikman, S. Nowak, \CQG{20}, 4799 (2003).
\ref [16] S. Mignemi, \MPL{A18}, 643 (2003).

\ref [18] S. Mignemi, \IJMP{D15}, 925 (2006).
\ref [19] S. Mignemi, in preparation.

\ref [21] F. Girelli, T. Konopka, J. Kowalski-Glikman, E.R. Livine,
\PR{D73}, 045009 (2006)
\endref
\end
Also, it is interesting to see what is the effect of the deformed
on the scattering of ultra-high photons by the cosmic background radiation [5]
As discussed in ref.\ [3] for any DSR model holds the approximate relation,
$$E_{th}=U^\mo({m_e^2\over E_{IR}})$$
where $m_e$ is the mass of the electron, $E_{IR}$ is the energy of the background photon,
and $E_{th}$ is the threshold energy. In our case, one obtains
$$E_{th}={m_e^2\over E_{IR}}\Big/\sqrt{1+{m_e^4\over E_{IR}^2}},$$

We also give a realization of the Snyder model in \des space\footnote{$^2$}{[17]}.
More details will be given in a forthcoming paper [12].
We consider the case $\O>0$, which, as in flat space, admits a maximal value $\k$
for the mass.
Starting from a 5-dimensional
space with \coo $\x_A$ satisfying the \des constraint $\x_A^2=-1/\L$, we define
$$x_\m=\sqrt{1+\O\p^2\over1+\L\x^2}\ \x_\m\qquad
p_\m=\sqrt{1+\L\x^2\over1+\O\p^2}\ \p_\m,\eqno(13)$$
The \coo (13) obey \pb
$$\eqalignno{
&\{x_\m,x_\n\}=-{\O\,(\lx)\over\lox}\ (x_\m p_\n-x_\n p_\m),&\cr
&\{p_\m,p_\n\}=-{\L\,(\op)\over\lox}\ (x_\m p_\n-x_\n p_\m),&\cr
&\{x_\m,p_\n\}=\y_\mn-{\L\,(\op)x_\m x_\n+\O\,(\lx)p_\m p_\n\over\lox}
\,,&(14)}$$
which become singular for $x^2p^2\to1/\L\O\sim10^{120}$...

The dilatation generators are
$$T_\m=\sqrt{\lox\over\op}\ p_\m,\eqno(15)$$
and their action is given by
$$\eqalignno{
&\{T_\m,x_\n\}=-\sqrt{\lox\over\op}\left[\y_\mn-{\L(\op)\over\lox}
\left(x_\m x_\n+\O\,{(\lx)\,x\pun p\,p_\m x_\n\over\lox}\right)\right],&\cr
&\{T_\m,p_\n\}=-\L\sqrt{\op\over\lox}\
\left[x_\m p_\n-x_\n p_\m+\O\,{(\lx)\,x\pun p\,p_\m p_\n\over\lox}\right].
&(16)}$$

The hamiltonian can be defined starting from the Casimir operator of the \des algebra
and reads [12]
$$H=\ha\left[{\lx\over\op}\ p^2+\L(x\pun p)^2\right].\eqno(17)$$

One can also define a metric invariant under the action of the \des group, starting
from the 5-dimensional metric on the \des hyperboloid,
$$g_\mn={\op\over\lx}\left[\y_\mn+\L{1+\O(\lx)p^2\over(\lx)(\lox)}\,x_\m x_\n
\right].\eqno(18)$$

In the limit $\L\to0$ one of course recovers the flat-space Snyder model,
while in the limit $\O\to0$ one gets the standard \des space, although with
noncanonical \pb between positions and momenta (since the momenta are identified
with the translation generators in this limit).
More interesting are the limits $x\to\a$ and $p\to\k$. For $x\to\a$, one is
close to the cosmological horizon, and the symplectic structure reduces to the
undeformed one obtained in the limit $\O=0$.
The limit $p\to\k$ corresponds instead to the extremal value of the momentum.
In this limit, the symplectic structure is that of the flat Snyder model, $\L=0$,
and the metric and the hamiltonian are singular.

For negative $\L$, one obtains the analogous of \ads space, with no cosmological horizon.
Also noticeable that for $\L$ and $\O$ both negative

Another interesting property of this model is the existence of a duality for the
exchange of $x\leftrightarrow p$, together with $\L\leftrightarrow\O$.
This duality connects the high-energy/short-distance regime, governed by the Planck
area $\O$, with the low-energy/long-distance regime, governed by the cosmological
constant $\L$.